\documentclass[aps,twocolumn,showpacs,showkeys]{revtex4}

\usepackage{amsmath}

\begin{document}


\def\beq{\begin{equation}}
\def\eeq{\end{equation}}
\def\bea{\begin{eqnarray}}
\def\eea{\end{eqnarray}}
\def\ket#1{|#1\rangle}
\def\bra#1{\langle#1|}


\title{\Large\textbf{Secret parameters in quantum bit commitment}}

\author{\textbf{Chi-Yee Cheung}}
\email{cheung@phys.sinica.edu.tw}

\affiliation{Institute of Physics, Academia Sinica\\
             Taipei 11529, Taiwan, Republic of China\\}


\begin{abstract}

The no-go theorem of unconditionally secure quantum bit
commitment depends crucially on the assumption that Alice
knows in detail all the probability distributions generated
by Bob. We show that if a protocol is concealing, then the
cheating unitary transformation is independent of any
parameters (including probability distributions) secretly
chosen by Bob, so that Alice can calculate it without
knowing Bob's secret choices. Otherwise the protocol cannot
be concealing. Our result shows that the original
impossibility proof was based on an incorrect assumption,
despite the fact that its conclusion remains valid within
the adopted framework. Furthermore, our result eliminates a
potential loophole in the no-go theorem.

\end{abstract}

\pacs{03.67.Dd, 03.67.Hk, 03.67.Mn}

\keywords{quantum bit commitment, quantum cryptography}

\maketitle


The security of quantum bit commitment (QBC) is an
important issue in quantum cryptography because QBC is a
primitive which can be used as the building block of other
important two-party cryptographic protocols \cite{BC96}.

A QBC protocol involves two parties customarily named Alice
and Bob. Alice secretly commits to a bit $b$ (0 or 1) which
is to be revealed to Bob at a later time. In order to bind
Alice to her commitment, the two parties execute a series
of quantum and/or classical procedures, so that at the end
of the commitment phase, Bob is in possession of a quantum
mechanical state $\ket{\psi^{(b)}_{B}}$. The idea is that,
with additional classical information from Alice in the
unveiling phase (when she unveils the value of $b$), Bob
can use $\ket{\psi^{(b)}_{B}}$ to check whether Alice is
honest. A QBC protocol is said to be binding if Alice
cannot change her commitment or Bob will find out.
Furthermore it is concealing if Bob can obtain no
information about the value of $b$ before it is unveiled,
which implies that the encoding density matrix
$\rho_B^{(b)}$ of the state $\ket{\psi^{(b)}_{B}}$ is
independent of the value of $b$, i.e.,
 \beq
 \rho_B^{(0)}=\rho_B^{(1)}\label{rho01}.
 \eeq
A QBC protocol is secure if and only if it is both binding
and concealing.  Moreover, if a protocol is secure even if
Alice and Bob had unlimited computational power, then it is
said to be unconditionally secure.

In 1997, Lo and Chau \cite{LoChau1,LoChau2} and Mayers
\cite{Mayers1,Mayers2} proved that unconditionally secure
QBC is impossible.  In a nutshell, the proof goes as
follows. It is observed that the commitment process, which
may involves any number of rounds of quantum and classical
exchange of information between Alice and Bob, can always
be represented by an unitary transformation ${\cal
U}^{(b)}$ on some initial state $\ket{\phi_{AB}}$ in the
combined Hilbert space $H_A\otimes H_B$ of Alice and Bob:
 \beq
 \ket{\Psi^{(b)}_{AB}}={\cal U}^{(b)}\ket{\phi^{(b)}_{AB}},
 \eeq
Without loss of generality, we can take
$\ket{\phi^{(b)}_{AB}}$ and $\ket{\Psi^{(b)}_{AB}}$ to be
pure states. In this approach, Alice and Bob do not fix
their undisclosed classical parameters in the commitment
phase, but leave them undetermined at the quantum level
instead. This is called quantum purification. In general it
requires that Alice and Bob have access to quantum
computers with unrestricted capacities, which is consistent
with the assumption that they have unlimited computational
power.

Therefore instead of honestly following the original
protocol, Alice can always follow a modified protocol as
described above, so that at the end of the commitment
phase, there exists a pure state $\ket{\Psi^{(b)}_{AB}}$ in
$H_A\otimes H_B$. As long as the reduced density matrix on
Bob's side is unchanged, i.e.,
 \beq
 {\rm Tr}_A~ \ket{\Psi^{(b)}_{AB}}
 \bra{\Psi^{(b)}_{AB}}
 =\rho_B^{(b)},
 \eeq
Bob has no way of knowing what Alice has actually done.
Then it follows from Schmidt decomposition theorem
\cite{HJW,Schmidt} that,
 \beq
 \ket{\Psi^{(0)}_{AB}}=\sum_i \sqrt{\lambda^i}\,
 \ket{e^i_A}\otimes\ket{\psi^i_B},
 \eeq
and
 \beq
 \ket{\Psi^{(1)}_{AB}}=\sum_i \sqrt{\lambda^i}\,
 \ket{e'^i_A}\otimes\ket{\psi^i_B},
 \eeq
where $\{\ket{e^i_A}\}$, $\{\ket{e'^i_A}\}$, and
$\{\ket{\psi^i_B}\}$ are orthonormal bases in the
respective Hilbert spaces as indicated, and $\lambda^i$'s
are real coefficients.

Notice that apart from the sets of bases $\{\ket{e^i_A}\}$
and $\{\ket{e'^i_A}\}$, $\ket{\Psi^{(0)}_{AB}}$ and
$\ket{\Psi^{(1)}_{AB}}$ are identical. Since
$\{\ket{e^i_A}\}$ and $\{\ket{e'^i_A}\}$ are related by an
unitary transformation $U_A$ acting on Alice's Hilbert
space $H_A$ only, we also have
 \beq
 \ket{\Psi^{(1)}_{AB}} = U_A \ket{\Psi^{(0)}_{AB}}.
 \label{UA}
 \eeq
The existence of $U_A$ implies that Alice has a sure-win
cheating strategy (called EPR attack): Alice always commits
to $b=0$ in the beginning.  Later on, if she wants to keep
her initial commitment, she unveils as prescribed. However
if she wants to switch to $b=1$ instead, she just needs to
apply the unitary transformation $U_A$ to the particles in
her control, and then proceeds as if she had committed to
$b=1$ in the first place. The crucial point is that,
because of Eq. (\ref{rho01}), it is impossible for Bob to
find out what Alice actually did, and he would conclude
that she is honest in either case. Hence if a QBC protocol
is concealing, it cannot be binding at the same time.  This
is the conclusion of the ``no-go theorem" of
unconditionally secure QBC.

Note that the no-go theorem only proves the existence of
the cheating unitary transformation $U_A$ in a QBC protocol
which is concealing, but there is no proof that $U_A$ is
always known to Alice. The point is, at the end of the
commitment phase, the overall state
$\ket{\Psi^{(b)}_{AB}(\omega)}$ may depend on some unknown
parameter $\omega$ secretly chosen by Bob. If the reduced
density matrix
 \beq
 \rho^{(b)}_B(\omega)={\rm Tr}_A
 \ket{\Psi^{(b)}_{AB}(\omega)}
 \bra{\Psi^{(b)}_{AB}(\omega)}
 \eeq
is independent of $b$, then in principle a cheating
transformation $U_A(\omega)$ exists, so that
 \beq
 \ket{\Psi^{(1)}_{AB}(\omega)}=U_A(\omega)
 \ket{\Psi^{(0)}_{AB}(\omega)}.
 \eeq
However without the knowledge of $\omega$, Alice cannot
calculate $U_A(\omega)$ by herself. As a result
unconditionally secure QBC may be possible. This is a
potential loophole of the no-go theorem.

The no-go theorem emphasizes that one should purify all
undisclosed classical variables in analyzing the security
issues. Even so, the question remains: What if Bob is
allowed to choose probability distributions secretly? To
this question, the authors of the no-go theorem state that
``In order that Alice and Bob can follow the procedures,
they must know the exact forms of all unitary
transformations involved" \cite{LoChau1,LoChau2}, and ``It
is a principle that we must assume that every participant
knows every detail of the protocol, including the
distribution of probability of a random variable generated
by another participant" \cite{Mayers2}. In other words the
no-go theorem asserts without proof that in any QBC
protocol the overall state $\ket{\Psi^{(b)}_{AB}}$ cannot
contain any unknown parameters. This assertion is in fact
not correct, and it has caused confusion among researchers.
Without clarifying this issue, the impossibility proof is
not complete and the no-go theorem will continue being
challenged \cite{Yuen}. In any case, as long as it does not
jeopardize the security of a protocol, there is no reason
why a party has to disclose the values of any secret
parameters he/she might have chosen in the commitment
phase.

To settle this issue, we prove the following theorem. The
secret parameter $\omega$ will be taken to be a probability
distribution, because in a fully quantum description,
probability distributions are the only unknowns left.
Except for the issue of secret parameters, we shall stay
within the QBC framework adopted by the no-go theorem.

\vspace{0.5em}

\textbf{Theorem 1}~~If a QBC protocol is concealing, then
the cheating unitary transformation is independent of any
probability distributions ($\omega$'s) secretly chosen by
Bob.

\vspace{0.5em}

\textbf{Proof}~~If Bob is allowed to choose $\omega$ in
secret, he can always postpone his choice with the help of
a quantum computer. That means, instead of picking a
particular $\omega=\omega_i$ and keeping it secret, he can
purify his choices with a probability distribution
$\pi=\{p_i\}$. The resulting overall state is given by
 \beq
 \ket{\Psi'^{(b)}_{AB}(\pi)}=\sum_i\sqrt{p_i}\,
 \ket{\Psi^{(b)}_{AB}(\omega_i)}~\ket{\chi_i},
 \label{Psi'}
 \eeq
where $\{\ket{\chi_i}\}$ is a set of orthonormal ancilla
states in Bob's Hilbert space $H_B$.
The new density matrix is given by
 \beq
 \rho'^{(b)}_B(\pi)={\rm Tr}_A
 \ket{\Psi'^{(b)}_{AB}(\pi)}\bra{\Psi'^{(b)}_{AB}(\pi)}.
 \eeq
Since the protocol is concealing, we have
 \beq
 \rho'^{(0)}_B(\pi)=\rho'^{(1)}_B(\pi)  \label{rho'2}
 \eeq
for all possible $\pi$. Consider the case where $p_i\ne 0$,
for all $i$. According to the no-go theorem there exists a
cheating unitary transformation $U'_A$, such that
 \beq
 \ket{\Psi'^{(1)}_{AB}(\pi)}=
 U'_A~\ket{\Psi'^{(0)}_{AB}(\pi)}
 \label{UA'1}.
 \eeq
It is easy to see that this same $U'_A$ also transforms
$\ket{\Psi^{(0)}_{AB}(\omega_i)}$ to
$\ket{\Psi^{(1)}_{AB}(\omega_i)}$ for all possible
$\omega_i$, \textit{i.e.},
 \beq
 \ket{\Psi^{(1)}_{AB}(\omega_i)}
 =U'_A~\ket{\Psi^{(0)}_{AB}(\omega_i)},
 \quad \forall \omega_i.
 \label{UA'2}
 \eeq
The reason is that Bob can obtain
$\ket{\Psi^{(b)}_{AB}(\omega_i)}$ from
$\ket{\Psi'^{(b)}_{AB}(\pi)}$ by collapsing the ancilla
states $\{\ket{\chi_i}\}$ on the right hand side of Eq.
(\ref{Psi'}). Since $U'_A$ acts on Alice's Hilbert space
$H_A$ only, it commutes with any operations executed on
Bob's Hilbert space $H_B$. Consequently Eq. (\ref{UA'1})
holds independent of whether the ancilla has been measured
or not, and Eq. (\ref{UA'2}) follows. Hence $U'_A$ is
independent of $\omega$.

To avoid a circular argument, we need to show that $U'_A$
is also independent of the probability distribution
$\pi=\{p_i\}$. As shown in the Appendix, any superposition
of probability distributions can be rewritten as a single
effective distribution. That is, by a redefinition of the
ancilla states on Bob's side, we can rewrite
$\ket{\Psi'^{(b)}_{AB}(\pi)}$ of Eq. (\ref{Psi'}) as
 \beq
 \ket{\Psi'^{(b)}_{AB}(\pi)}=
 \ket{\Psi^{(b)}_{AB}(\omega_j(\pi))}.
 \label{Psi''}
 \eeq
Substituting Eq. (\ref{Psi''}) into Eq. (\ref{UA'1}), we
see that $U'_A$ might depend on $\pi$ through
$\omega_j(\pi)$. But that is not possible since we have
already proved that $U'_A$ is independent of $\omega_i$ for
all $i$ [see Eq. (\ref{UA'2})]. Hence the EPR cheating
transformation $U'_A$ does not dependent on $\pi$. QED.

Therefore in a concealing QBC protocol, the cheating
unitary transformation $U_A$ is independent of any secret
probability distributions chosen by Bob, and Alice can
calculate $U_A$ without knowing Bob's particular choices.
In fact, according to the corollary proven in the Appendix,
$U_A$ cannot depend on any probability distribution
(specified or secret) generated by Bob. This contradicts
the claim that, to be able to cheat, Alice must know every
detail of the protocol, including all the probability
distributions generated by Bob, so that no unknown
parameter is allowed in $\ket{\Psi^{(b)}_{AB}}$
\cite{LoChau1,LoChau2,Mayers1,Mayers2}.

Conversely, if in any protocol the cheating unitary
transformation is claimed to depend on a secret parameter
$\omega$ chosen by Bob, then the protocol must be
non-concealing under closer scrutiny. Thus our result
eliminates a potential loophole in the no-go theorem.

In summary, we find that there is nothing wrong with secret
parameters in QBC. We prove that in a concealing protocol,
the cheating unitary transformation is independent of any
parameters secretly chosen by Bob. Our result shows that
the original proof of the no-go theorem
\cite{LoChau1,LoChau2,Mayers1,Mayers2} was based on an
incorrect assumption. Nevertheless, even with secret
parameters, unconditional security remains impossible
within the framework adopted by the no-go theorem.


\vskip 0.5truecm

\centerline{\textbf{Appendix}}

\vskip 0.5truecm

Suppose Bob chooses an unitary operator in $\{V_k\}$ with a
probability distribution $\omega_i=\{q_{ik}\}$ and applies
it to a state $\ket{\phi_{AB}}$, such that
 \beq
 \ket{\psi_{AB}}
 = V_k \ket{\phi_{AB}}.
 \eeq
As is well known, if $V_k$ is not disclosed, Bob can
postpone (or purify) his decision by entangling with a set
of orthonormal ancilla states $\{\ket{\xi_k}\}$, so that
instead of $\ket{\psi_{AB}}$, he generates
 \beq
 \ket{\Psi_{AB}(\omega_i)}=
 \sum_k \sqrt{q_{ik}}\, \ket{\xi_k} V_k
 \ket{\phi_{AB}}.
 \label{entangle1}
 \eeq
Likewise, if $\omega_i$ is not disclosed, Bob can also
purify his choices with another probability distribution
$\pi=\{p_i\}$, such that
 \bea
 \ket{\Psi'_{AB}(\pi)}&=&
 \sum_{i} \sqrt{p_i}\, \ket{\chi_i}
 \ket{\Psi_{AB}(\omega_i)},\label{entangle2}\\*
 &=&\sum_{i,k} \sqrt{p_i}\, \sqrt{q_{ik}}\,
 \ket{\chi_i} \ket{\xi_k} V_k
 \ket{\phi_{AB}},
 \label{entangle3}
 \eea
where $\{\ket{\chi_i}\}$ is another set of orthonormal
ancilla states.

\vspace{0.5em}

\textbf{Theorem 2}~~Purifying a probability distribution
[in Eq. (\ref{entangle2})] is equivalent to picking a new
effective one [in Eq. (\ref{entangle1})].

\vspace{0.5em}

\textbf{Proof}~~Define
 \beq
 q'_k\equiv\sum_i ~p_i q_{ik},
 \eeq
so that
 \beq
 \sum_k q'_k=1.
 \eeq
On the right hand side of Eq. (\ref{entangle3}), we write
 \bea
 \sum_i \sqrt{p_i q_{ik}}\, \ket{\chi_i}
 &=&\sqrt{q'_k}\, \sum_i \sqrt{p_i q_{ik}/q'_k}\,
 \ket{\chi_i}\nonumber\\
 &=&\sqrt{q'_k}\, \ket{\chi'_k}, \label{qchi}
 \eea
where
 \beq
 \ket{\chi'_k}\equiv\sum_i\sqrt{p_i q_{ik}/q'_k}\,
 \ket{\chi_i}.
 \eeq
Note that the $\ket{\chi'_k}$'s are normalized but not
necessarily orthogonal.

Substituting Eq. (\ref{qchi}) into Eq. (\ref{entangle3}),
we get
 \bea
 \ket{\Psi'_{AB}(\pi)} \!\!&=&\!\!
 \sum_k \sqrt{q'_k}\, \ket{\chi'_k}\ket{\xi_k}V_k
 \ket{\phi_{AB}},\nonumber\\
 \!\!&=&\!\!\sum_k \sqrt{q'_k}\, \ket{\xi'_k}V_k
 \ket{\phi_{AB}},
 \eea
where
 \beq
 \ket{\xi'_k}\equiv \ket{\chi'_k}\ket{\xi_k}
 \eeq
are the new orthonormal ancilla states. Comparing with Eq.
(\ref{entangle1}), we obtain the desired result
 \beq
 \ket{\Psi'_{AB}(\pi)} =
 \ket{\Psi_{AB}(\omega_j(\pi))},
 \label{effective}
 \eeq
where $\omega_j(\pi)\equiv\{q'_k\}$ is the new effective
probability distribution.

Using the above result, we can prove the following
corollary:

\vspace{0.5em}

\textbf{Corollary}~~It is in general not meaningful to
specify a probability distribution to an untrustful party
in any quantum protocol, because he/she can always cheat.

\vspace{0.5em}

\textbf{Proof}~~Suppose the protocol specifies that Bob
should take certain action $V_k$ on each qubit (or group of
qubits) in his possession according to a probability
distribution $\omega_j=\{q'_k\}$. According to the theorem
just proven, he can always generate a superposition of
distributions with appropriately chosen $p_i$'s, such that
the effective distribution is $\omega_j$ [see Eqs.
(\ref{entangle2}, \ref{effective})]. Obviously Bob would
have no problem passing any checks concerning $\omega_j$.
In general some qubits are measured and discarded in the
checking procedure. For each of the remaining qubits, Bob
could either stay with $\omega_j$, or he could measure the
ancilla states $\{\ket{\chi_i}\}$ in Eq. (\ref{entangle2})
to obtain a new distribution $\omega_i$ which is not
necessarily equal to $\omega_j$. For a large number of
qubits, the probability that $\omega_j$ is obtained for
every qubit is exponentially small. Hence it is not
meaningful to specify a probability distribution to an
untrustful Bob, because one can never be sure that he is
honest.

\vskip 1.0truecm

\acknowledgments  It is a pleasure to thank H. P. Yuen, H.
-K. Lo, C. H. Bennett, and S. Popescu for helpful
discussions. This work is supported in part by National
Science Council of the Republic of China under Grant NSC
92-2112-M-001-058.


\end{document}